\documentclass[12pt,aps,prd,preprint,tightenlines,superscriptaddress,showpacs,nofootinbib]
{revtex4-1}
\usepackage{epsfig}
\usepackage{amssymb,amsmath}

\newcommand{\bea}{\begin{eqnarray}}
\newcommand{\eea}{\end{eqnarray}}

 \makeatletter
\def\alt{\mathrel{\mathpalette\gl@align<}}
\def\agt{\mathrel{\mathpalette\gl@align>}}
\def\gl@align#1#2{\lower.6ex\vbox{\baselineskip\z@skip\lineskip\z@
\ialign{$\m@th#1\hfil##\hfil$\crcr#2\crcr\sim\crcr}}} \makeatother

\begin{document}
\begin{flushright}
YGHP16-06
\end{flushright}
\vspace*{1.0cm}

\begin{center}
\baselineskip 20pt 
{\Large\bf 
$Z^\prime$-portal right-handed neutrino dark matter\\
 in the minimal U(1)$_X$ extended Standard Model 
}
\vspace{1cm}

{\large 
Nobuchika Okada$^{~a}$  and  Satomi Okada$^{~b}$
}
\vspace{.5cm}

{\baselineskip 20pt \it
$^a$Department of Physics and Astronomy, University of Alabama, Tuscaloosa, AL35487, USA\\
$^b$Graduate School of Science and Engineering, Yamagata University,  \\
Yamagata 990-8560, Japan} 

\vspace{.5cm}

\vspace{1.5cm} {\bf Abstract}
\end{center}

We consider a concise dark matter (DM) scenario in the context of a non-exotic U(1) extension 
  of the Standard Model (SM), where a new U(1)$_X$ gauge symmetry is introduced along with 
  three generation of right-handed neutrinos (RHNs) and an SM gauge singlet Higgs field.  
The model is a generalization of the minimal gauged U(1)$_{B-L}$ (baryon number minus lepton number) extension of the SM, 
  in which the extra U(1)$_X$ gauge symmetry is expressed as a linear combination 
  of the SM U(1)$_Y$ and U(1)$_{B-L}$ gauge symmetries. 
We introduce a $Z_2$-parity and assign an odd-parity only for one RHN among all particles, 
  so that this $Z_2$-odd RHN plays a role of DM. 
The so-called minimal seesaw mechanism is implemented in this model 
  with only two $Z_2$-even RHNs. 
In this context, we investigate physics of the RHN DM, 
  focusing on the case that this DM particle communicates 
  with the SM particles through the U(1)$_X$ gauge boson ($Z^\prime$ boson). 
This ``$Z^\prime$-portal RHN DM" scenario is controlled by only three free parameters: 
  the U(1)$_X$ gauge coupling ($\alpha_X$), the $Z^\prime$ boson mass ($m_{Z^\prime}$), 
  and the U(1)$_X$ charge of the SM Higgs doublet ($x_H$).  
We consider various phenomenological constraints to identify a phenomenologically viable parameter space. 
The most important constraints are the observed DM relic abundance and 
  the latest LHC Run-2 results on the search for a narrow resonance with the di-lepton final state.  
We find that these are complementary with each other and narrow the allowed parameter region,  
  leading to the lower mass bound of $m_{Z^\prime} \gtrsim 2.7$ TeV. 

\thispagestyle{empty}

\newpage

\addtocounter{page}{-1}

\baselineskip 18pt

\section{Introduction} 
\label{sec:1}

Neutrino masses and a suitable candidate for the dark matter are the major missing pieces in the Standard Model (SM), 
  which require us to extend the SM. 
The minimal $B-L$ model~\cite{Mohapatra:1980qe, Marshak:1979fm, Wetterich:1981bx, Masiero:1982fi, Mohapatra:1982xz,Buchmuller:1991ce} is a simple, well-motivated extension of the SM to incorporate the neutrino masses, 
  where the global U(1)$_{B-L}$ (baryon number minus lepton number) symmetry in the SM is gauged. 
In the presence of the three right-handed neutrinos (RHNs) the model is free from all the gauge and gravitational anomalies. 
Associated with the spontaneous $B-L$ gauge symmetry breaking by a vacuum expectation value (VEV) 
 of the $B-L$ Higgs field, 
 the RHNs and the $B-L$ gauge boson ($Z^\prime$ boson) acquire their masses.  
With the generated Majorana masses for the RHNs, 
  the seesaw mechanism~\cite{Minkowski:1977sc,  Yanagida:1979as, GellMann:1980vs, Glashow:1979nm, Mohapatra:1979ia} 
  is implemented, and the light SM neutrino mass is generated after the electroweak symmetry breaking. 
The mass spectrum of the new particles introduced in the minimal $B-L$ model 
   (the $Z^\prime$ boson, the Majorana RHNs and the $B-L$ Higgs boson)  
  is controlled by the $B-L$ gauge symmetry breaking scale. 
If the breaking scale lies around the TeV scale, the $B-L$ model can be tested 
  at the Large Hadron Collider (LHC).

Among various possibilities, a concise way of introducing a dark matter (DM) candidate in the minimal $B-L$ model 
   has been proposed in Ref.~\cite{Okada:2010wd}. 
Instead of extending the minimal particle content, a $Z_2$-parity is introduced 
   and an odd-parity is assigned to only one RHN  
  while even-parities are assigned to all the other fields.\footnote{
We can consider the $Z_2$-parity as an emergent global symmetry 
  in the limit of vanishing Dirac Yukawa couplings~\cite{Anisimov:2008gg}.
}  
Hence, the parity-odd RHN serves as the DM. 
On the other hand, two parity-even RHNs account for the neutrino mass generation via the seesaw mechanism. 
This system is nothing but the so-called minimal seesaw~\cite{King:1999mb, Frampton:2002qc}, 
  which is the minimal setup to reproduce the observed neutrino oscillation data 
  with a prediction of one massless neutrino as well as the observed baryon asymmetry 
  in the universe through leptogenesis~\cite{Fukugita:1986hr}.

There are two ways for the RHN DM to communicate with the SM particles. 
One is through two Higgs bosons, which are expressed as linear combinations 
   of the SM Higgs and the $B-L$ Higgs bosons after the breaking 
   of the U(1)$_{B-L}$ and the electroweak gauge symmetries. 
The DM phenomenology for this case has been analyzed in \cite{Okada:2010wd, Okada:2012sg, Basak:2013cga}.    
The other way is that the interactions between the DM and the SM particles 
  are mediated by the $Z^\prime$ boson. 
This class of DM scenario is called  ``$Z^\prime$-portal DM" and 
   has been attracting a lot of attention recently. 
In the scenario, a DM particle is introduced along with an electric-charge neutral vector field 
   ($Z^\prime$ boson)  in an extension of the SM 
   with the so-called Dark Sector~\cite{An:2012va, An:2012ue,  Soper:2014ska, Sierra:2015fma} 
   or new gauge interactions~\cite{ 
Burell:2011wh, Basso:2012ti, Das:2013jca,  Chu:2013jja, Dudas:2013sia, Lindner:2013awa, Alves:2013tqa, Kopp:2014tsa, 
Agrawal:2014ufa,Hooper:2014fda, Ma:2014qra, Alves:2015pea,  
Ghorbani:2015baa, Sanchez-Vega:2015qva, 
Duerr:2015wfa, Alves:2015mua,Ma:2015mjd, Okada:2016gsh, Okada:2016tzi, Chao:2016avy, Biswas:2016ewm, 
Accomando:2016sge, Fairbairn:2016iuf, Kaneta:2016vkq, Klasen:2016qux, Dev:2016xcp, Altmannshofer:2016jzy}. 
The mediator $Z^\prime$ boson allows us to investigate a variety of DM physics, 
  such as the DM relic abundance and the direct/indirect DM search.  
A remarkable feature of the scenario is that the $Z^\prime$ boson resonance search  
  at the LHC is complementary to the cosmological observations of the $Z^\prime$-portal DM 
  in identifying a phenomenologically viable parameter region.

Recently, the minimal $B-L$ model with the RHN DM has been investigated 
  in the light of the LHC Run-2 results~\cite{Okada:2016gsh}.   
Here, the RHN DM communicates with the SM particles mainly 
  through the $Z^\prime$ gauge boson, and hence it is the $Z^\prime$-portal DM scenario. 
In the model, the DM physics is controlled by 
  only two free parameters, the $B-L$ gauge coupling and the $Z^\prime$ boson mass. 
It has been found that the constraint from the observed DM relic abundance 
  leads to a lower bound on the gauge coupling as a function of the $Z^\prime$ boson mass. 
On the other hand, the cross section of $Z^\prime$ boson production at the LHC 
   is also determined by the same two free parameters. 
The LHC  Run-2 results on search for a narrow resonance with the di-lepton final states 
   have been interpreted to obtain the upper bound on the gauge coupling as a function of the $Z^\prime$ boson mass. 
Combining the two results, an allowed parameter region has been identified 
   to obtain the lower bound of $m_{Z^\prime} \gtrsim 2.5$ TeV. 
In deriving the allowed parameter region, a complementarity 
  between the cosmological and the collider constraints was essential. 
     
In this paper, we generalize the minimal $B-L$ model to the so-called non-exotic 
   U(1)$_X$ extension of the SM~\cite{Appelquist:2002mw}.  
The  non-exotic U(1)$_X$ model is the most general extension of the SM with an extra 
   anomaly-free U(1) gauge symmetry, which is described as a linear  
   combination of the SM U(1)$_Y$ and the U(1)$_{B-L}$ gauge groups. 
The particle content of the model is the same as the one in the minimal $B-L$ model 
   except for the generalization of the U(1)$_X$ charge assignment for particles. 
Hence we can easily extend the minimal $B-L$ model with the RHN DM 
   to the non-exotic U(1)$_X$ case. 
In this context, we perform detailed analysis to identify a phenomenologically viable parameter region 
   through the complementarity between the  DM physics and the LHC Run-2 results. 
Because of  the U(1)$_X$ generalization, the $Z^\prime$ boson couplings 
  with the SM particles are modified and the resultant parameter region 
  is found to be quite different from the one obtained in Ref.~\cite{Okada:2016gsh}.  
For the LHC Run-2 results, we employ the most recent results reported by 
  the ATLAS and the CMS collaborations in 2016~\cite{ATLAS:2016, CMS:2016}.

This paper is organized as follows. 
In the next section, we define the minimal non-exotic U(1)$_X$ extension of the SM 
  with the $Z^\prime$-portal RHN DM. 
In Sec.~\ref{sec:3}, we analyze the DM relic abundance and identify a model parameter region 
  to satisfy the observed DM relic abundance. 
In Sec.~\ref{sec:4}, we consider the results by the ATLAS and the CMS collaborations 
  at the LHC Run-2 on the search for a narrow resonance with the di-lepton final states. 
We interpret the results into the constraints on the $Z^\prime$ boson production 
  in the minimal non-exotic U(1)$_X$ model.  
Combining all the constraints, we identify the allowed parameter regions in Sec.~\ref{sec:5}. 
The last section is devoted to conclusions.

\section{The minimal non-exotic U(1)$_X$ model with RHN DM}
\label{sec:2}
\begin{table}[t]
\begin{center}
\begin{tabular}{c|ccc|c|c}
      &  SU(3)$_c$  & SU(2)$_L$ & U(1)$_Y$ & U(1)$_X$  & $ Z_2 $\\ 
\hline
$q^{i}_{L}$ & {\bf 3 }    &  {\bf 2}         & $ 1/6$       & $(1/6) x_{H} + (1/3) x_{\Phi}$    & $+$\\
$u^{i}_{R}$ & {\bf 3 }    &  {\bf 1}         & $ 2/3$       & $(2/3) x_{H} + (1/3) x_{\Phi}$   & $+$\\
$d^{i}_{R}$ & {\bf 3 }    &  {\bf 1}         & $-1/3$       & $-(1/3) x_{H} + (1/3) x_{\Phi}$   &$+$\\
\hline
$\ell^{i}_{L}$ & {\bf 1 }    &  {\bf 2}         & $-1/2$       & $(-1/2) x_{H} - x_{\Phi}$  & $+$ \\
$e^{i}_{R}$    & {\bf 1 }    &  {\bf 1}         & $-1$                   & $-x_{H} - x_{\Phi}$   & $+$ \\
\hline
$H$            & {\bf 1 }    &  {\bf 2}         & $- 1/2$       & $(-1/2) x_{H}$  & $+$ \\  
\hline
$N^{j}_{R}$    & {\bf 1 }    &  {\bf 1}         &$0$                    & $- x_{\Phi}$   & $+$ \\
$N_{R}$         & {\bf 1 }    &  {\bf 1}         &$0$                    & $- x_{\Phi}$   & $-$ \\
$\Phi$            & {\bf 1 }       &  {\bf 1}       &$ 0$                  & $ + 2x_{\Phi}$  & $+$ \\ 
\hline
\end{tabular}
\end{center}
\caption{
The particle content of the minimal U(1)$_X$ extended SM with $Z_2$ parity. 
In addition to the SM particle content ($i=1,2,3$), the three RHNs 
  ($N_R^j$ ($j=1,2$) and $N_R$) and the U(1)$_X$ Higgs field ($\Phi$) are introduced.   
Because of the $Z_2$ parity assignment shown here, the $N_R$ is a unique (cold) DM candidate. 
The extra U(1)$_X$  gauge group is defined with a linear combination of the SM U(1)$_Y$   
  and the U(1)$_{B-L}$ gauge groups, and the U(1)$_X$ charges of fields are determined by 
  two real parameters, $x_H$ and $x_\Phi$.   
Without loss of generality, we fix $x_\Phi=1$ throughout this paper. 
}
\label{table1}
\end{table}

We first define our model by the particle content listed on Table~\ref{table1}. 
The U(1)$_X$ gauge group is identified with a linear combination of 
   the SM U(1)$_Y$  and the U(1)$_{B-L}$ gauge groups, 
   and hence the U(1)$_X$ charges of fields are determined by 
   two real parameters, $x_H$ and $x_\Phi$.   
Note that in the model the charge $x_\Phi$ always appears as a product 
   with the U(1)$_X$ gauge coupling and it is not an independent free parameter. 
Hence, we fix $x_\Phi=1$ throughout this paper. 
In this way, we reproduce the minimal $B-L$ model with the conventional charge assignment 
   as the limit of $x_H \to 0$. 
The limit of $x_H \to +\infty~(-\infty)$ indicates that the U(1)$_X$ is (anti-)aligned to the U(1)$_Y$ direction.  
The anomaly structure of the model is the same as the minimal $B-L$ model 
  and the model is free from all the gauge and the gravitational anomalies 
  in the presence of the three RHNs.  
The introduction of the $Z_2$-parity is crucial to incorporate a DM candidate in the model 
  while keeping the minimality of the particle content. 
The conservation of the $Z_2$-parity ensures the stability of the $Z_2$-odd RHN, 
  and therefore it is a unique DM candidate in the model.

The Yukawa sector of the SM is extended to have 
\bea
\mathcal{L}_{Yukawa} \supset  - \sum_{i=1}^{3} \sum_{j=1}^{2} Y^{ij}_{D} \overline{\ell^i_{L}} H N_R^j 
          -\frac{1}{2} \sum_{k=1}^{2} Y^k_N \Phi \overline{N_R^{k~C}} N_R^k 
          -\frac{1}{2}  Y_N \Phi \overline{N_R^{~C}} N_R 
  + {\rm h.c.} ,
\label{Lag1} 
\eea
where the first term is the neutrino Dirac Yukawa coupling, and the second and 
   third terms are the Majorana Yukawa couplings. 
Without loss of generality, the Majorana Yukawa couplings are already diagonalized in our basis.  
Note that because of the $Z_2$-parity, only the two generation RHNs are involved 
  in the neutrino Dirac Yukawa coupling. 
Once the U(1)$_X$ Higgs field $\Phi$ develops a nonzero VEV,  
   the U(1)$_X$ gauge symmetry is broken and the Majorana mass terms 
   for the RHNs are generated. 
Then, the seesaw mechanism is automatically implemented in the model after the electroweak symmetry breaking.  
Because of the $Z_2$-parity, only two generation RHNs are relevant to the seesaw mechanism. 
Even with two RHNs, the Yukawa coupling constants $Y_D^{ij}$ and $Y_N^k$ posses the degrees of freedom 
   large enough to reproduce the neutrino oscillation data with a prediction of one massless eigenstate.  
The baryon asymmetry in the universe can also be reproduced with the two RHNs~\cite{Frampton:2002qc}
(see, for example, Ref.~\cite{Iso:2010mv} for detailed analysis of leptogenesis at the TeV scale 
   with two RHNs).

The renormalizable scalar potential for the SM Higgs doublet ($H$) and the U(1)$_X$ Higgs fields is given by 
\bea  
V = \lambda_H \left(  H^{\dagger}H -\frac{v^2}{2} \right)^2
+ \lambda_{\Phi} \left(  \Phi^{\dagger} \Phi -\frac{v_X^2}{2}  \right)^2
+ \lambda_{\rm mix} 
\left(  H^{\dagger}H -\frac{v^2}{2} \right) 
\left(  \Phi^{\dagger} \Phi -\frac{v_X^2}{2}  \right) , 
\label{Higgs_Potential }
\eea
where all quartic couplings are chosen to be positive. 
At the potential minimum, the Higgs fields develop their VEVs as 
\bea
  \langle H \rangle =  \left(  \begin{array}{c}  
    \frac{v}{\sqrt{2}} \\
    0 \end{array}
\right),  \;  \;  \; \; 
\langle \Phi \rangle =  \frac{v_X}{\sqrt{2}}. 
\eea
In this paper, we assume $\lambda_{\rm mix} \ll1$, so that the mixing 
   between the SM Higgs boson and the U(1)$_X$ Higgs boson are negligibly small.\footnote{
This assumption is, in fact, not essential. 
When $\lambda_{\rm mix}$ is sizable, the RHN DM can communicate with the SM particles 
  also through the Higgs bosons. 
This so-called Higgs portal RHN DM case has been analyzed in  \cite{Okada:2010wd, Okada:2012sg, Basak:2013cga} 
  and it has been shown that the RHN DM mass is required to be close to a half of either one of the Higgs boson masses 
  in order to reproduce the observed relic abundance. 
Such a parameter region is distinguishable from that in our $Z^\prime$-portal RHN DM case, 
  and we can investigate the two cases separately.  
}   
Hence, the RHN DM communicates with the SM particles only through the $Z^\prime$ boson. 
Associated with the U(1)$_X$ symmetry breaking, the Majorana neutrinos $N_R^j$ $(j=1,2)$, 
  the DM particle $N_R$ and the $Z^\prime$ gauge boson acquire their masses as 
\bea 
  m_N^j=\frac{Y_N^j}{\sqrt{2}} v_X,  \; \; 
  m_{DM}=\frac{Y_N}{\sqrt{2}} v_X,  \; \; 
  m_{Z^\prime} = g_X \sqrt{4 v_X^2+  \frac{v^2}{4}} \simeq 2 g_X v_X , 
\eea  
where $g_X$ is the U(1)$_X$ gauge coupling, 
   and we have used the LEP constraint~\cite{LEP:2003, LEP:2013} $v_X^2 \gg v^2$. 
Because of the LEP constraint, the mass mixing of the $Z^\prime$ boson with the SM $Z$ boson
   is very small, and we neglect it in our analysis in this paper.

Assuming $\lambda_{\rm mix} \ll 1$, we focus on the $Z^\prime$-portal nature of the RHN DM. 
In this case, only four free parameters ($g_X$, $m_{Z^\prime}$, $m_{DM}$, and $x_H$) 
   are involved in our analysis. 
As we will discuss in the next section, it turns out that the condition of $m_{DM} \simeq m_{Z^\prime}/2$  
   must be satisfied to reproduce the observed DM relic abundance.
Thus, $m_{DM}$ does not work as an independent parameter, 
  so that our results are described by only three free  parameters.

\section{Cosmological constraints on $Z^\prime$-portal RHN DM.}
\label{sec:3}
In the Planck satellite experiments, the DM relic abundance is measured 
   at the 68\% limit as \cite{Aghanim:2015xee}  
\bea 
   \Omega_{DM} h^2 = 0.1198\pm 0.0015.  
\label{Planck_DM}
\eea 
In this section, we evaluate the DM relic abundance and identify an allowed parameter region 
   to satisfy the upper bound of $\Omega_{DM} h^2 \leq 0.1213$.  
The DM relic abundance is evaluated by integrating the Boltzmann equation given by 
\bea 
  \frac{dY}{dx}
  = - \frac{x s \langle \sigma v \rangle}{H(m_{DM})} \left( Y^2-Y_{EQ}^2 \right), 
\label{Boltmann}
\eea  
where the temperature of the universe is normalized by the mass of the RHN DM as $x=m_{DM}/T$, 
   $H(m_{DM})$ is the Hubble parameter at $T=m_{DM}$, 
   $Y$ is the yield (the ratio of the DM number density to the entropy density $s$) of 
  the RHN DM, $Y_{EQ}$ is the yield of the DM particle in thermal equilibrium, 
  and $\langle \sigma v \rangle$ is the thermal average of the DM annihilation cross section times relative velocity ($v$). 
Explicit formulas of the quantities involved in the Boltzmann equation are as follows: 
\bea 
s = \frac{2  \pi^2}{45} g_\star \frac{m_{DM}^3}{x^3} ,  \; \;
H(m_{DM}) =  \sqrt{\frac{\pi^2}{90} g_\star} \frac{m_{DM}^2}{M_P},  \; \;
s Y_{EQ}= \frac{g_{DM}}{2 \pi^2} \frac{m_{DM}^3}{x} K_2(x),   
\eea
where $M_P=2.44 \times 10^{18}$  GeV is the reduced Planck mass, 
   $g_{DM}=2$ is the number of degrees of freedom for the DM particle, 
   $g_\star$ is the effective total number of degrees of freedom for the particles in thermal equilibrium 
   (in the following analysis, we use $g_\star=106.75$ for the SM particles),  
   and $K_2$ is the modified Bessel function of the second kind.   
In our $Z^\prime$-portal DM scenario, a DM pair annihilates into the SM particles 
    through the $Z^\prime$ boson exchange in the $s$-channel.  
The thermal average of the annihilation cross section is given by 
\bea 
\langle \sigma v \rangle = \left(s Y_{EQ} \right)^{-2} g_{DM}^2
  \frac{m_{DM}}{64 \pi^4 x} 
  \int_{4 m_{DM}^2}^\infty  ds \; \hat{\sigma}(s) \sqrt{s} K_1 \left(\frac{x \sqrt{s}}{m_{DM}}\right) , 
\label{ThAvgSigma}
\eea
where $\hat{\sigma}(s)=2 (s- 4 m_{DM}^2) \sigma(s)$ is the reduced cross section 
   with the total annihilation cross section $\sigma(s)$, and $K_1$ is the modified Bessel function of the first kind. 
The total cross section of the DM pair annihilation process $NN \to Z^\prime \to f {\bar f}$ 
   ($f$ denotes the SM fermions) is calculated as 
\bea 
 \sigma(s)=\frac{\pi}{3}  \alpha_X^2  \frac{\sqrt{s (s-4 m_{DM}^2)}}
  {(s-m_{Z^\prime}^2)^2+m_{Z^\prime}^2 \Gamma_{Z^\prime}^2} 
    F(x_H),  
\label{DMSigma}
\eea 
where 
\bea 
  F(x_H)=13+ 16 x_H  + 10 x_H^2 = 10  \left( x_H+0.8 \right)^2 +6.6, 
\label{F}  
\eea
  and the total decay width of $Z^\prime$ boson is given by 
\bea
\Gamma_{Z'} = 
 \frac{\alpha_X}{6} m_{Z^\prime} 
 \left[ F(x_H) + \left( 1-\frac{4 m_{DM}^2}{m_{Z^\prime}^2} \right)^{\frac{3}{2}} 
 \theta \left( \frac{m_{Z^\prime}^2}{m_{DM}^2} - 4 \right)  \right]. 
\label{width}
\eea
Here, we have neglected all SM fermion masses and assumed $m_N^j > m_{Z^\prime}/2$, for simplicity.

\begin{figure}[t]
\begin{center}
\includegraphics[width=0.465\textwidth,angle=0,scale=1.05]{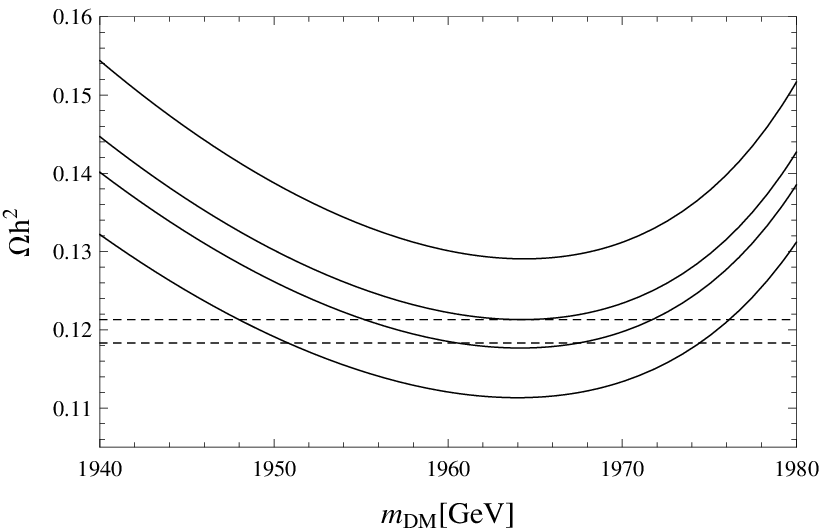}
\hspace{0.1cm}
\includegraphics[width=0.465\textwidth,angle=0,scale=1.05]{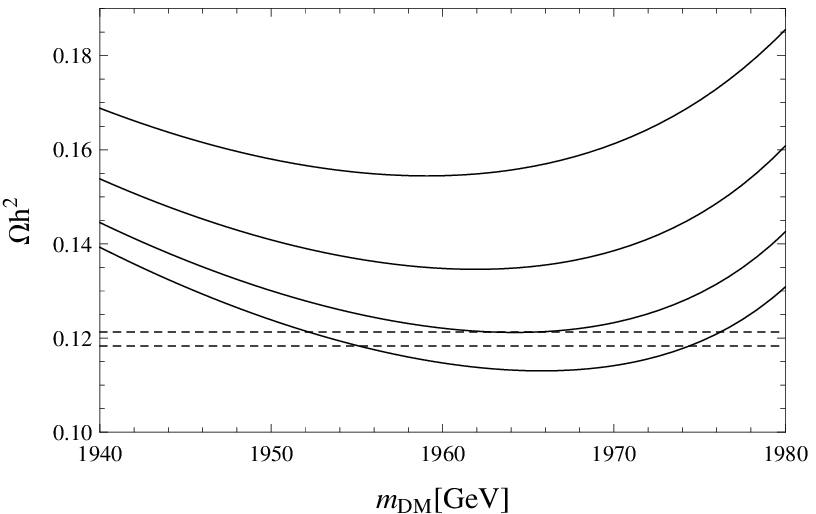}
\end{center}
\caption{
The relic abundance of the $Z^\prime$-portal RHN DM 
  as a function of its mass ($m_{DM}$) for $m_{Z^\prime}=4$ TeV. 
In the left panel, we have fixed $x_H=0$ (the minimal $B-L$ model limit) and shown 
  the relic abundance for various values of the gauge coupling,  
  $\alpha_X=0.025$, $0.027$, $0.028$ and $0.030$ 
  (solid lines from top to bottom). 
In the right panel, we have fixed $\alpha_X=0.027$ and shown 
  the relic abundance for various values of 
  $x_H = -0.8$, $0$, $0.5$ and $1.0$ 
  (solid lines from bottom to top). 
The two horizontal lines denote the range of the observed DM relic density, 
  $0.1183 \leq \Omega_{DM} h^2 \leq 0.1213$ in Eq.~(\ref{Planck_DM}).
}
\label{fig:Omega}
\end{figure}

\begin{figure}[t]
\begin{center}
\includegraphics[width=0.6\textwidth,angle=0,scale=1.06]{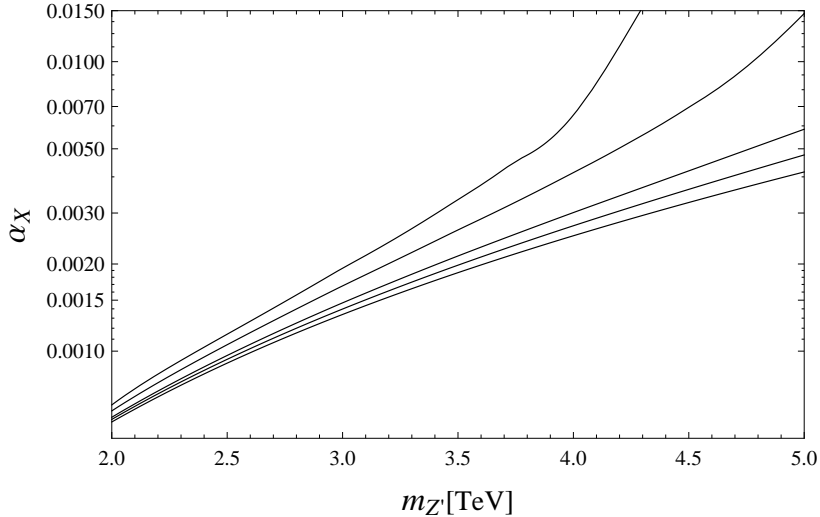} 
\end{center}
\caption{
The lower bounds on $\alpha_X$ as a function of $m_{Z^\prime}$ 
  for various values of $x_H$, 
  to satisfy the cosmological constraint of $0.1183 \leq \Omega_{DM} h^2 \leq 0.1213$.   
The solid lines from top to bottom corresponds to 
  $x_H=-3$, $+1$, $-2$, $0$ and $-1$, respectively. 
As the input $x_H$ value is going away from the point of $x_H=-0.8$, 
  the lower bound on $\alpha_X$ is increasing. 
}
\label{fig:DMlines}
\end{figure}

Now we solve the Boltzmann equation numerically, and 
   find the asymptotic value of the yield $Y(\infty)$ to evaluate the present DM relic density as 
\bea 
  \Omega_{DM} h^2 =\frac{m_{DM} s_0 Y(\infty)} {\rho_c/h^2}, 
\eea 
  where $s_0 = 2890$ cm$^{-3}$ is the entropy density of the present universe, 
  and $\rho_c/h^2 =1.05 \times 10^{-5}$ GeV/cm$^3$ is the critical density.
Our analysis involves four parameters, 
   namely $\alpha_X=g_X^2/(4 \pi)$, $m_{Z^\prime}$, $m_{DM}$ and $x_H$.       
For $m_{Z^\prime}=4$ TeV, we show in Fig.~\ref{fig:Omega} the resultant DM relic abundance   
   as a function of the DM mass, 
  along with the range of the observed DM relic abundance, 
   $0.1183 \leq \Omega_{DM} h^2 \leq 0.1213$~\cite{Aghanim:2015xee} 
  (two horizontal dashed lines). 
In the left panel, we have fixed $x_H=0$, which is the minimal $B-L$ model limit.  
The solid lines from top to bottom show the resultant DM relic abundances
  for various values of the gauge coupling, $\alpha_X=0.025$, $0.027$, $0.028$ and $0.030$.  
The plots indicate the lower bound on $\alpha_X \geq 0.027$  
   for $m_{Z^\prime}=4$ TeV and $x_H=0$ 
   in order to be able to reproduce the observed relic abundance. 
In addition, we can see that the enhancement of the DM annihilation cross section 
   via the $Z^\prime$ boson resonance is necessary to satisfy the cosmological constraint 
   and hence, $m_{DM} \simeq m_{Z^\prime}/2$. 
The right panel shows our results for various values of $x_H$ with the fixed $\alpha_X=0.027$. 
The solid lines from bottom to top correspond to the results 
   for $x_H = -0.8$, $0$, $0.5$ and $1.0$, respectively. 
 From Eqs.~(\ref{ThAvgSigma})-(\ref{width}), we can see that the DM annihilation cross section 
   for $m_{DM} \simeq m_{Z^\prime}/2$ is proportional to  $1/F(x_H)$. 
Therefore, the maximum annihilation cross section for the fixed values of $\alpha_X$, $m_{Z^\prime}$ 
   and $m_{DM} \simeq m_{Z^\prime}/2$ is achieved for $x_H = -0.8$. 
Since the function $F(x_H)$ is symmetric about the point of  $x_H=-0.8$, 
   the results shown in the left panel indicate the constraint $-1.6 \leq  x_H \leq  0$ 
   to satisfy the cosmological bound 
   for the fixed values of $m_{Z^\prime}=4$ TeV and $\alpha_X = 0.027$.

In Fig.~\ref{fig:DMlines} we show the lower bounds on $\alpha_X$ as a function of $m_{Z^\prime}$ 
  for various values of $x_H$, to reproduce the observed DM relic abundance 
   in the range of $0.1183 \leq \Omega_{DM} h^2 \leq 0.1213$. 
The solid lines from top to bottom corresponds to 
  $x_H=-3$, $+1$, $-2$, $0$ and $-1$, respectively. 
For fixed $\alpha_X$ and $m_{Z^\prime}$,  the DM annihilation cross section 
  becomes maximum for $x_H \simeq -0.8$ with the minimum $Z^\prime$ boson decay width. 
As an input $x_H$ value is going away from the point of $x_H=-0.8$, 
  the decay width becomes larger and the DM annihilation cross section is reducing. 
As a result, the lower bound on the gauge coupling is increasing.

\section{LHC Run-2 constraints}  
\label{sec:4}
In 2015, the LHC Run-2 started its operation with a 13 TeV collider energy. 
The most recent results by the ATLAS and the CMS collaborations with the combined 2015 and 2016 data    
  were reported at the ICHEP 2016 conference. 
The ATLAS and the CMS collaborations continue their search for $Z^\prime$ boson resonance 
  with di-lepton final states at the LHC Run-2. 
Their results have shown significant improvements 
  for the upper limits of the $Z^\prime$ boson production cross section~\cite{ATLAS:2016, CMS:2016} 
  from those obtained by the LHC Run-1~\cite{ATLAS8TeV, CMS8TeV}.     
In this section, we will employ the most recent LHC Run-2 results to derive 
  LHC constraints on the model parameters, $\alpha_{X}$, $m_{Z^\prime}$ and $x_H$. 

Let us calculate the cross section for the process $pp \to Z^\prime +X \to \ell^{+} \ell^{-} +X$. 
The differential cross section with respect to the invariant mass $M_{\ell \ell}$ of the final state di-lepton 
   is given by
\begin{eqnarray}
 \frac{d \sigma}{d M_{\ell \ell}}
 =  \sum_{q, {\bar q}}
 \int^1_ \frac{M_{\ell \ell}^2}{E_{\rm CM}^2} dx
 \frac{2 M_{\ell \ell}}{x E_{\rm CM}^2}  
 f_q(x, Q^2) f_{\bar q} \left( \frac{M_{\ell \ell}^2}{x E_{\rm CM}^2}, Q^2
 \right)  {\hat \sigma} (q \bar{q} \to Z^\prime \to  \ell^+ \ell^-) ,
\label{CrossLHC}
\end{eqnarray}
where $f_q$ is the parton distribution function for a parton (quark) ``$q$'', 
  and $E_{\rm CM} =13$ TeV is the center-of-mass energy of the LHC Run-2.
In our numerical analysis, we employ CTEQ6L~\cite{CTEQ} for the parton distribution functions 
   with the factorization scale $Q= m_{Z^\prime}$. 
Here, the cross section for the colliding partons is given by 
\bea 
{\hat \sigma}(q \bar{q} \to Z^\prime \to  \ell^+ \ell^-) =
\frac{\pi}{1296} \alpha_X^2 
\frac{M_{\ell \ell}^2}{(M_{\ell \ell}^2-m_{Z^\prime}^2)^2 + m_{Z^\prime}^2 \Gamma_{Z^\prime}^2} 
F_{q \ell}(x_H),  
\label{CrossLHC2}
\eea
where the function $F_{q \ell}(x_H)$ is given by 
\bea
   F_{u \ell}(x_H) &=&  (8 + 20 x_H + 17 x_H^2)  (8 + 12 x_H + 5 x_H^2),   \nonumber \\
   F_{d \ell}(x_H) &=&  (8 - 4 x_H + 5 x_H^2) (8 + 12 x_H + 5 x_H^2) 
\label{Fql}
\eea
for $q$ being the up-type ($u$) and down-type ($d$) quarks, respectively.  
By integrating the differential cross section over a range of $M_{\ell \ell}$ set by the ATLAS and the CMS analysis, 
  respectively, we obtain the cross section to be compared with the upper bounds 
  obtained by the ATLAS and the CMS collaborations.

\begin{figure}[t]
\begin{center}
\includegraphics[width=0.45\textwidth,angle=0,scale=1.06]{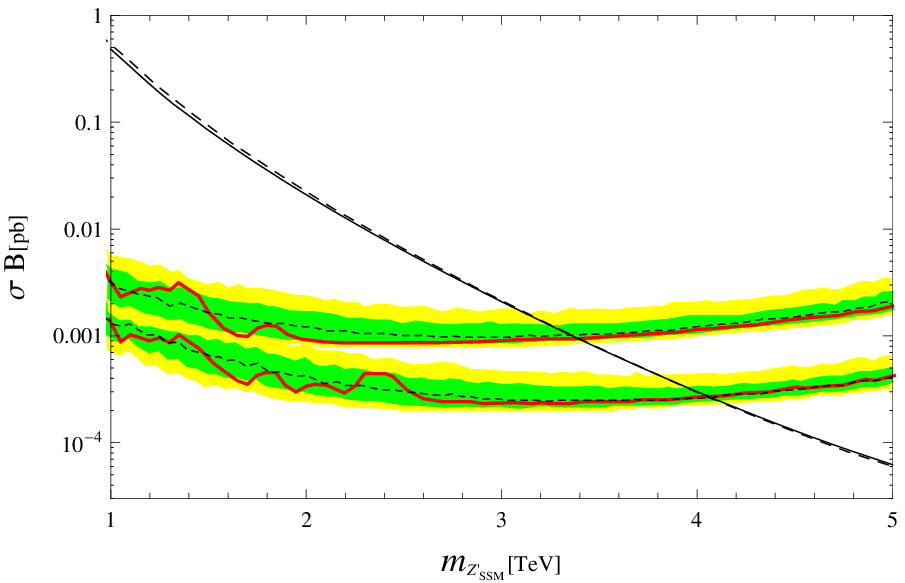} 
\hspace{0.1cm}
\includegraphics[width=0.45\textwidth,angle=0,scale=1.08]{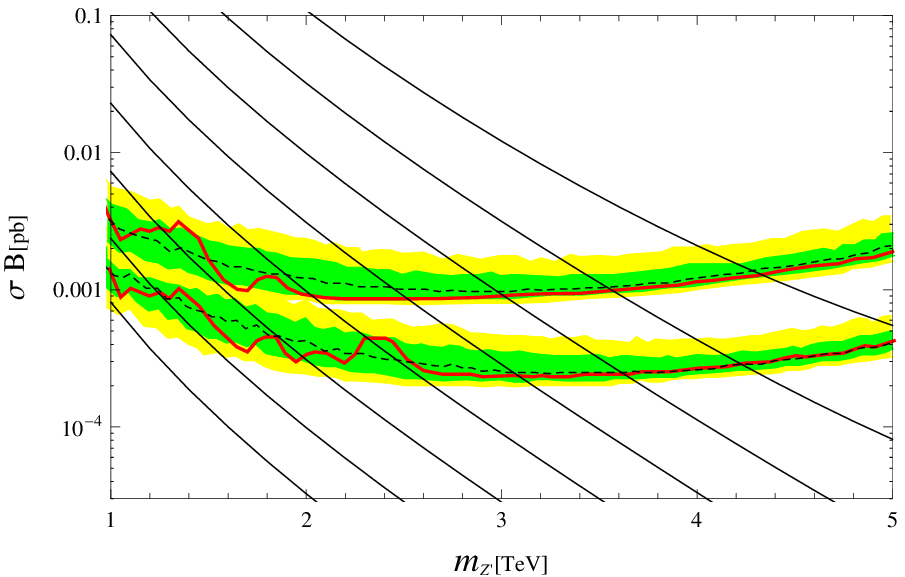}
\end{center}
\caption{
Left panel: the cross section as a function of the $Z^\prime_{SSM}$ mass (solid line) 
  with $k=1.28$, along with the ATLAS results in 2016~\cite{ATLAS:2016} and 2015~\cite{ATLAS:2015} 
  from the combined di-electron and di-muon channels. 
Right panel: the cross sections calculated for various values of $\alpha_X$ with $k=1.28$, 
   for the minimal $B-L$ model limit ($x_H=0$).
The solid lines from left to right correspond to 
   $\alpha_{X}=10^{-5}$,  $10^{-4.5}$, $10^{-4}$,  $10^{-3.5}$, $10^{-3}$, $10^{-2.5}$, $10^{-2}$, and $10^{-1.5}$,  
   respectively. 
}
\label{fig:Mssm-ATL}
\end{figure}

In the analysis by the ATLAS and the CMS collaborations, 
   the so-called sequential SM $Z^\prime$ ($Z^\prime_{SSM}$) model~\cite{ZpSSM} 
   has been considered as a reference model.  
We first analyze the sequential $Z^\prime$ model to check a consistency of our analysis 
   with the one by the ATLAS collaboration.  
In the sequential $Z^\prime$ model, the $Z^\prime_{SSM}$ boson has exactly the same 
   couplings with quarks and leptons as the SM $Z$ boson. 
With the couplings, we calculate the cross section of the process $pp \to Z^\prime_{SSM}+X \to \ell^+ \ell^- +X$ 
   like Eq.~(\ref{CrossLHC}). 
By integrating the differential cross section in the region of 128 GeV$\leq M_{\ell \ell} \leq 6000$ GeV~\cite{ATLAS8TeV}, 
  we obtain the cross section of the di-lepton production process as a function of $Z^\prime_{SSM}$ boson mass.
Our result is shown as a solid line in the left panel on Fig.~\ref{fig:Mssm-ATL}, 
  along with the plot presented by the ATLAS collaboration~\cite{ATLAS:2015, ATLAS:2016}.  
In the ATLAS paper~\cite{ATLAS:2016}, the lower limit of the $Z^\prime_{SSM}$ boson mass 
  is found to be $4.05$ TeV,  
   which is read off from the intersection point of the theoretical prediction (diagonal dashed line) and  
   the experimental cross section bound (lower horizontal solid curve (in red)).  
Here, we have also shown the plot presented in Ref.~\cite{ATLAS:2015} 
   (upper horizontal solid curve (in red)).  
We can see the dramatic improvement from the 2015 results~\cite{ATLAS:2015} 
   to the 2016 results~\cite{ATLAS:2016}.  
In order to take into account the difference of the parton distribution functions used in the ATLAS and our analysis 
  and QCD corrections of the process, we have scaled our resultant cross section by a factor $k=1.28$, 
  with which we can obtain the same lower limit of the $Z^\prime_{SSM}$ boson mass as $4.05$ TeV.  
We can see that our result with the factor of $k=1.28$ (solid line) is very consistent with the theoretical prediction 
  (diagonal dashed line) presented in Ref.~\cite{ATLAS:2016}. 
This factor is used in our analysis of the $Z^\prime$ boson production process in the following.

Now we calculate the cross section of the process  $pp \to Z^\prime+X \to \ell^+ \ell^- +X$ 
   for various values of $\alpha_X$, $m_{Z^\prime}$ and $x_H$.  
For $x_H=0$ (the minimal $B-L$ model limit), 
   we show our results in the right panel of Fig.~\ref{fig:Mssm-ATL}, 
   along with the plots in the ATLAS papers~\cite{ATLAS:2016,ATLAS:2015}. 
The diagonal solid lines from left to right correspond to 
   $\alpha_{X}=10^{-5}$,  $10^{-4.5}$, $10^{-4}$,  $10^{-3.5}$, 
    $10^{-3}$, $10^{-2.5}$, $10^{-2}$, and $10^{-1.5}$. 
From the intersections of the lower horizontal curve (in red) and diagonal solid lines, we can read off 
   the lower bounds on the $Z^\prime$ boson mass for the corresponding $\alpha_X$ values.  
For example, $m_{Z^\prime} > 3.1$ TeV for $\alpha_X=0.001$.   
In this way, we have obtained the upper bound on $\alpha_X$ as a function of the $Z^\prime$ boson mass. 
For various values of $x_H$ we do the same analysis and find the upper bound.

\begin{figure}[t]
\begin{center}
\includegraphics[width=0.45\textwidth,angle=0,scale=1.06]{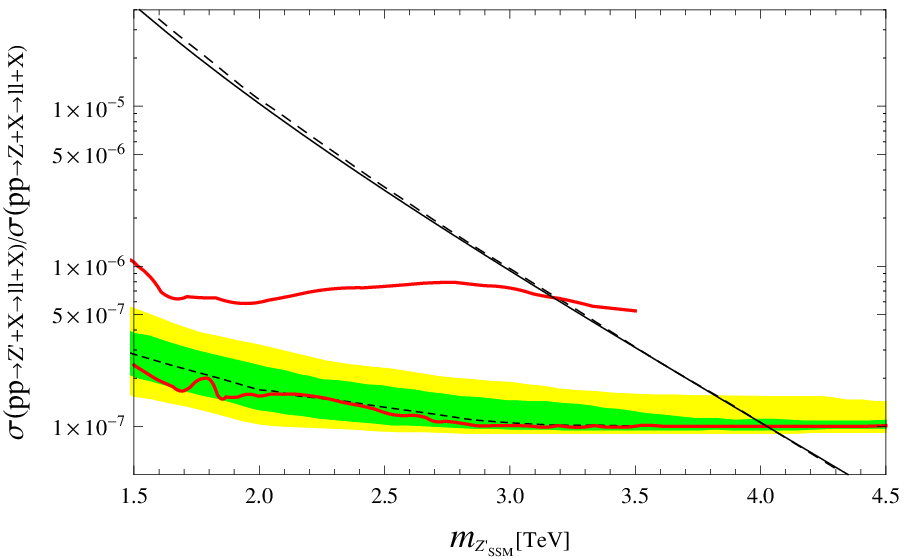} 
\hspace{0.1cm}
\includegraphics[width=0.45\textwidth,angle=0,scale=1.07]{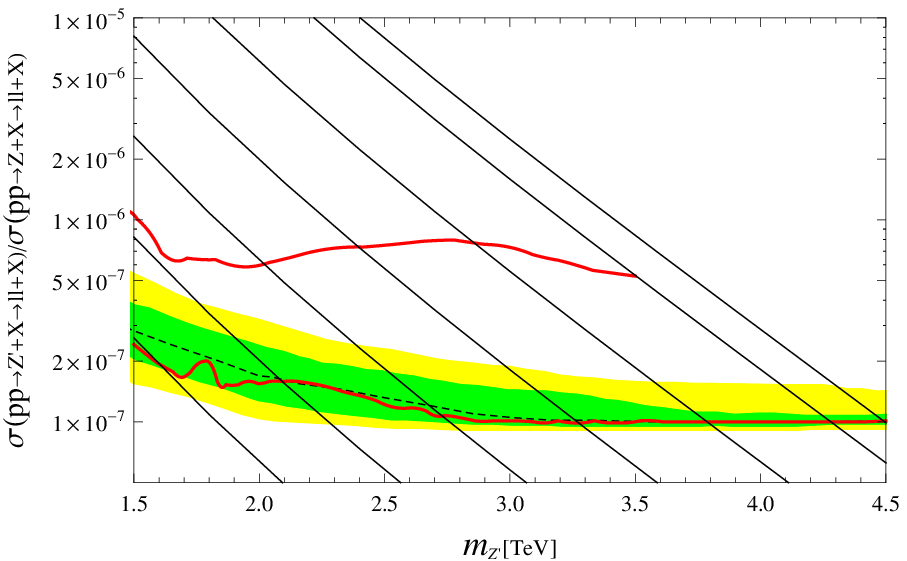}
\end{center}
\caption{
Left panel: the cross section ratio as a function of the $Z^\prime_{SSM}$ mass (solid line) 
  with $k=1.61$, along with the CMS results in 2015~\cite{CMS:2015} and 2016~\cite{CMS:2016}
   from the combined di-electron and di-muon channels. 
Right panel: the cross section ratios calculated for various values of $\alpha_X$ with $k=1.61$ 
    for $x_H=0$.
The solid lines from left to right correspond to 
   $\alpha_{X}=10^{-4.5}$, $10^{-4}$,  $10^{-3.5}$, $10^{-3}$, $10^{-2.5}$, $10^{-2}$, and $10^{-1.75}$,  
   respectively.  
}
\label{fig:Mssm-CMS}
\end{figure}

\begin{figure}[t]
\begin{center}
\includegraphics[width=0.6\textwidth,angle=0,scale=1.06]{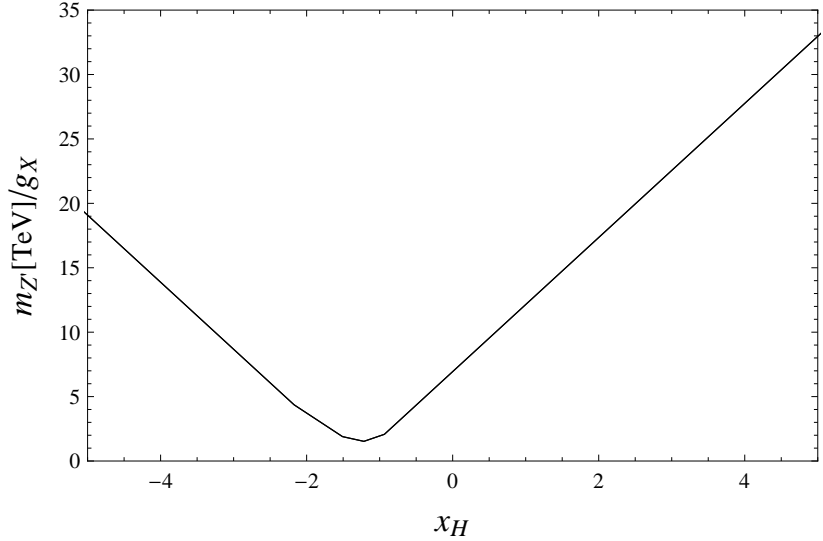} 
\end{center}
\caption{
The lower bound on $m_{Z^\prime}/g_X$ as a function of $x_H$. 
We have employed the final LEP 2 data~\cite{LEP:2013} at 95\% confidence level.   
}
\label{fig:LEP}
\end{figure}

\begin{figure}[t]
\begin{center}
\includegraphics[width=0.45\textwidth,angle=0,scale=1.06]{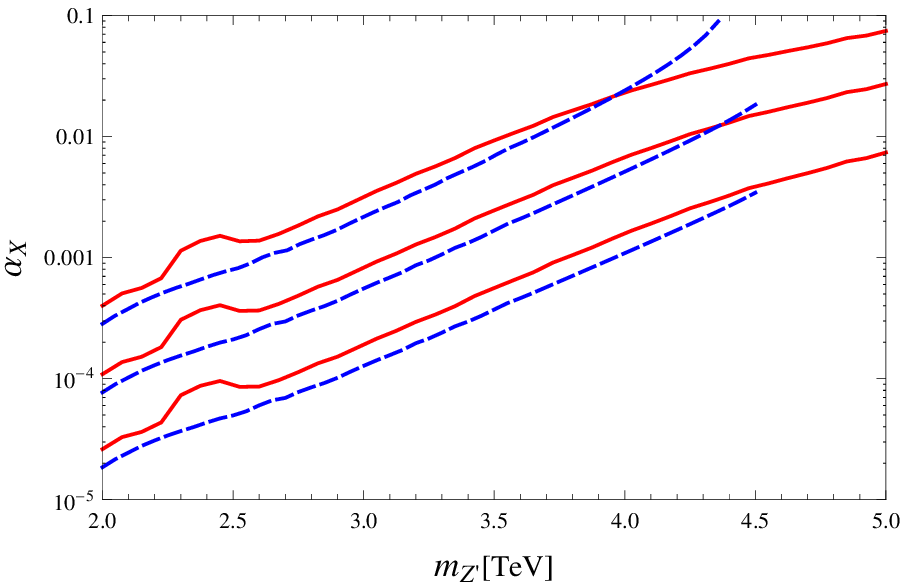} 
\hspace{0.1cm}
\includegraphics[width=0.45\textwidth,angle=0,scale=1.06]{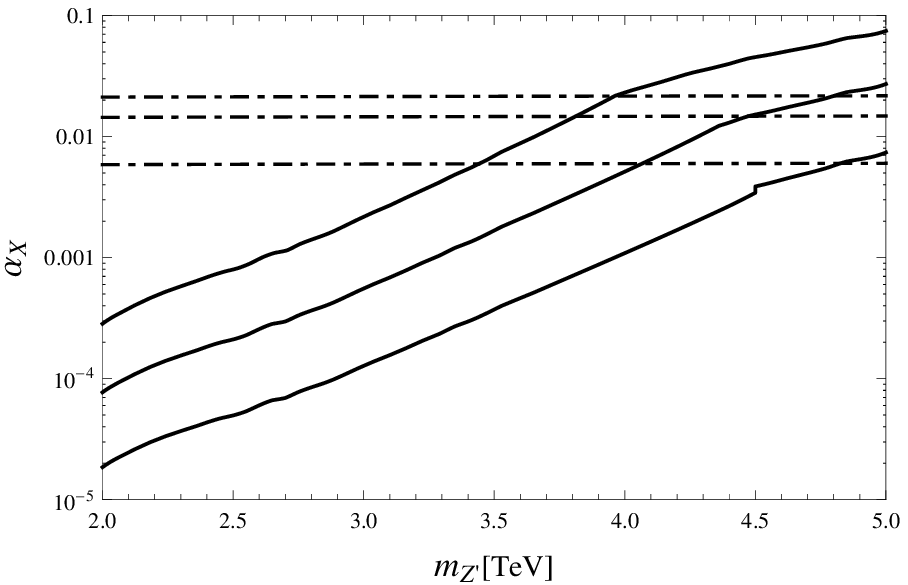}
\end{center}
\caption{
Left panel: the upper bounds on $\alpha_X$ as a function of $m_{Z^\prime}$ 
  for $x_H=-1$, $0$ and $+1$ from top to bottom for both of the solid and dashed lines, respectively.  
The solid lines denote the bounds from the ATLAS results~\cite{ATLAS:2016} 
  while the dashed lines denote the bounds from the CMS results~\cite{CMS:2016}. 
Right panel: the upper bounds on $\alpha_X$ after combining the ATLAS and the CMS 
  results shown in the left panel. 
The solid lines correspond to the combined upper bounds for $x_H=-1$, $0$ and $+1$ 
  from top to bottom, respectively.  
The perturbativity bounds of Eq.~(\ref{pert}) for $x_H=-1$, $0$ and $+1$ 
  are shown as the horizontal dashed-dotted lines from top to bottom, respectively. 
}
\label{fig:ATL-CMS}
\end{figure}

We apply the same strategy and compare our result for the $Z^\prime_{SSM}$ model 
   with the one by the CMS collaboration~\cite{CMS:2015, CMS:2016}.
According to the CMS analysis, we integrate the differential cross section in the range of 
  $0.95 \leq  M_{\ell \ell}/m_{Z^\prime_{SSM}} \leq  1.05$.  
In the CMS analysis, a limit has been set on the ratio of the $Z^\prime_{SSM}$ boson cross section 
  to the $Z/\gamma^*$ cross section in a mass window of 60 to 120 GeV, 
  which is predicted to be $1928$ pb.   
Our result is shown as a diagonal solid line in the left panel of Fig.~\ref{fig:Mssm-CMS}, 
  along with the plot presented in Ref.~\cite{CMS:2016}. 
The analysis in this CMS paper leads to the lower limit of the $Z^\prime_{SSM}$ boson mass as $4.0$ TeV,  
   which is read off from the intersection point of the theoretical prediction (diagonal dashed line) and  
   the experimental cross section bound (lower horizontal solid curve (in red)).  
Here, we have also shown the plot presented in Ref.~\cite{CMS:2015} 
   (upper horizontal solid curve (in red)).  
As in the left panel of Fig.~\ref{fig:Mssm-ATL}, we can see the dramatic improvement 
  from the 2015 results~\cite{CMS:2015} to the 2016 results~\cite{CMS:2016}.  
In order to obtain the same lower mass limit of $m_{Z^\prime_{SSM}} \geq 4.0$ TeV,  
   we have introduced a factor $k=1.61$.  
We can see that our result (solid line) are very consistent 
  with the theoretical cross section (dashed line) presented in Ref.~\cite{CMS:2016}.

With the factor of $k=1.61$, we have calculated the cross section of the process  $pp \to Z^\prime +X \to \ell^+ \ell^- +X$ 
   for various values of $\alpha_X$, $m_{Z^\prime}$ and $x_H$.  
For the minimal $B-L$ model limit, 
   we show our results in the right panel of Fig.~\ref{fig:Mssm-CMS}, 
   along with the plots in the CMS papers~\cite{CMS:2015, CMS:2016}. 
The diagonal solid lines from left to right correspond to 
   $\alpha_{X}=10^{-4.5}$, $10^{-4}$,  $10^{-3.5}$, 
    $10^{-3}$, $10^{-2.5}$, $10^{-2}$, and $10^{-1.75}$. 
From the intersections of the lower horizontal curve and the diagonal solid lines, we can read off 
   the lower bounds on the $Z^\prime$ boson mass for the corresponding $\alpha_X$ values.  
For example, $m_{Z^\prime} > 3.8$ TeV for $\alpha_X=10^{-2.5}$.   
In this way, we have obtained the upper bound on $\alpha_X$ as a function of the $Z^\prime$ boson mass. 
For various values of $x_H$ we do the same analysis and find the upper bound.

The search for effective 4-Fermi interactions mediated by a $Z^\prime$ boson at the LEP 
   leads to a lower bound on $m_{Z^\prime}/g_X$~\cite{LEP:2003, LEP:2013}. 
Employing the limits from the final LEP 2 data~\cite{LEP:2013} at 95\% confidence level, 
    we follow Ref.~\cite{Carena:2004xs} and derive a lower bound on $m_{Z^\prime}/g_X$
    as a function of $x_H$. 
Our result is shown in Fig.~\ref{fig:LEP}.   
For example, we find   
\bea
   \frac{m_{Z^\prime}}{g_{X}} \geq 6.94 \; {\rm TeV}. 
 \label{LEP}
\eea 
  for the minimal $B-L$ model limit, which is consistent with the result found in Ref.~\cite{Heeck:2014zfa}.   
We find that for any values of $x_H$, the LEP constraints are always weaker 
  than the LHC Run-2 constraints for $m_{Z^\prime} \leq 5$ TeV.

As a theoretical constraint, we may impose an upper bound on the U(1)$_X$ gauge coupling 
   to avoid the Landau pole in its renormalization group evolution $\alpha_{X}(\mu)$ 
  up to the Plank mass, $1/\alpha_{X}(M_{Pl}) > 0$, 
  where $M_{Pl}=1.22 \times 10^{19}$ GeV.  
Let us define the gauge coupling $\alpha_{X}$ used in our analysis 
  for the dark matter physics and LHC physics 
  as the running gauge coupling $\alpha_{X}(\mu)$ at $\mu=m_{Z^\prime}$. 
Employing the renormalization group equation at the one-loop level 
  with $m_N^1=m_N^2=m_\Phi=m_{Z^\prime}$, for simplicity, 
  we find 
\bea
  \alpha_{X} < \frac{2 \pi}{b_X \ln \left[ \frac{M_{Pl}}{m_{Z^\prime}} \right]}, 
  \label{pert}
\eea 
where $b_X= (72 + 64 x_H + 41 x_H^2)/6$ is the beta function coefficient. 
 
In Fig.~\ref{fig:ATL-CMS} we show the LHC Run-2 bounds 
  on $\alpha_X$ as a function of $m_{Z^\prime}$ for $x_H=-1$, $0$ and $+1$.  
In the left panel,  the solid (dashed) lines from top to bottom denote the upper bounds on $\alpha_X$
  for $x_H=-1$, $0$ and $+1$, respectively, obtained from the ATLAS results~\cite{ATLAS:2016} 
  (the CMS results~\cite{CMS:2016}). 
For $m_{Z^\prime} \lesssim 4-4.5$ TeV, the CMS bounds are slightly more severe than 
   those from the ATALS results. 
Combining the ATLAS and the CMS results, we obtain the upper bound 
  shown in the right panel.  
The solid lines corresponds to the combined upper bounds for $x_H=-1$, $0$ and $+1$ 
  from top to bottom, respectively.  
The perturbativity bounds of Eq.~(\ref{pert}) for $x_H=-1$, $0$ and $+1$ 
  are shown as the horizontal dashed-dotted lines from top to bottom, respectively.

\begin{figure}[t]
\begin{center}
\includegraphics[width=0.45\textwidth,angle=0,scale=1.06]{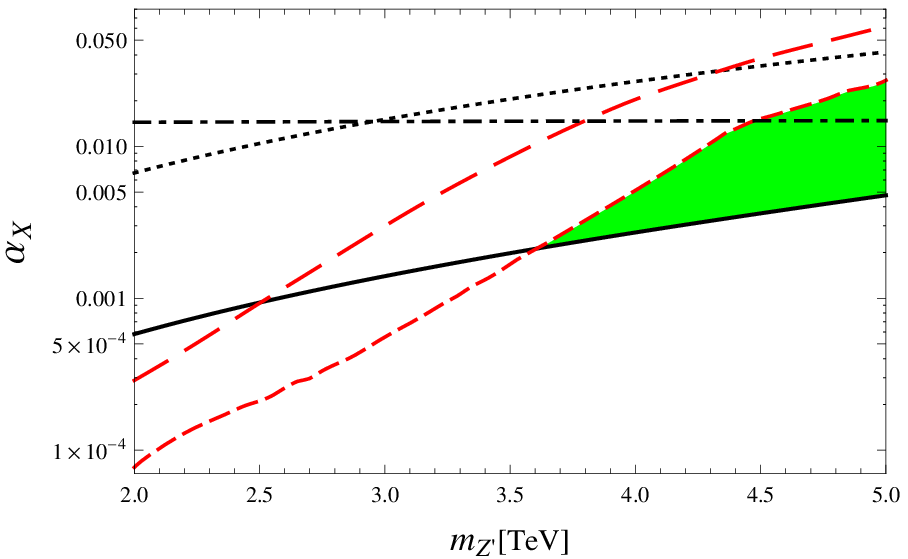} 
\hspace{0.1cm}
\includegraphics[width=0.45\textwidth,angle=0,scale=1.06]{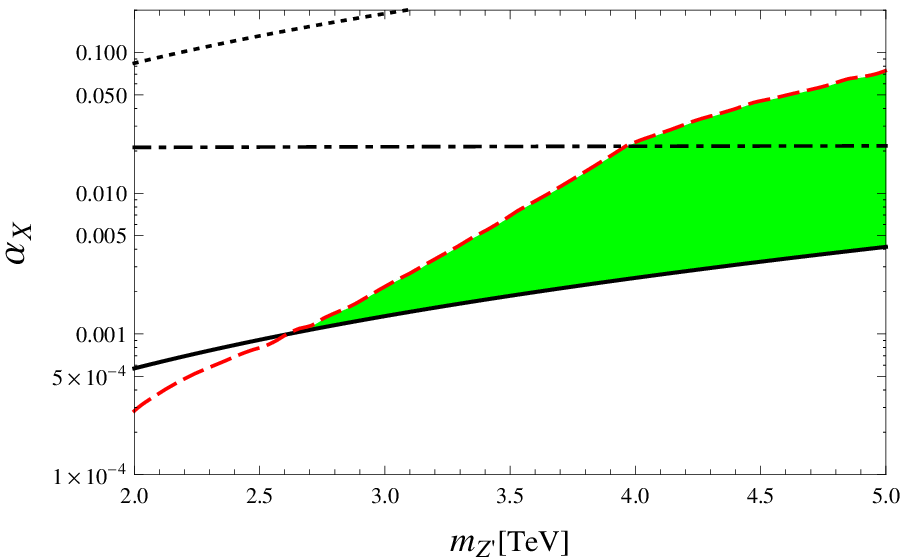}
\includegraphics[width=0.45\textwidth,angle=0,scale=1.06]{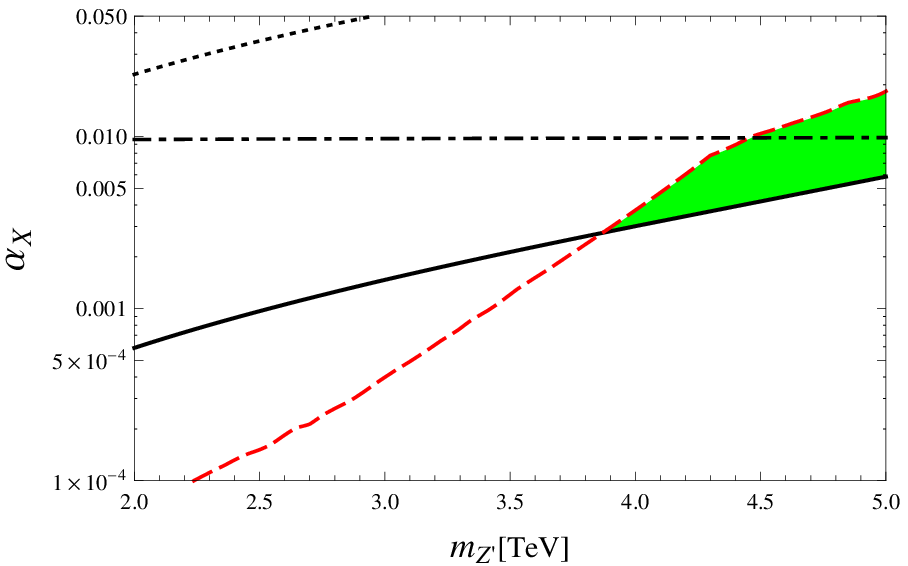}
\hspace{0.1cm}
\includegraphics[width=0.45\textwidth,angle=0,scale=1.06]{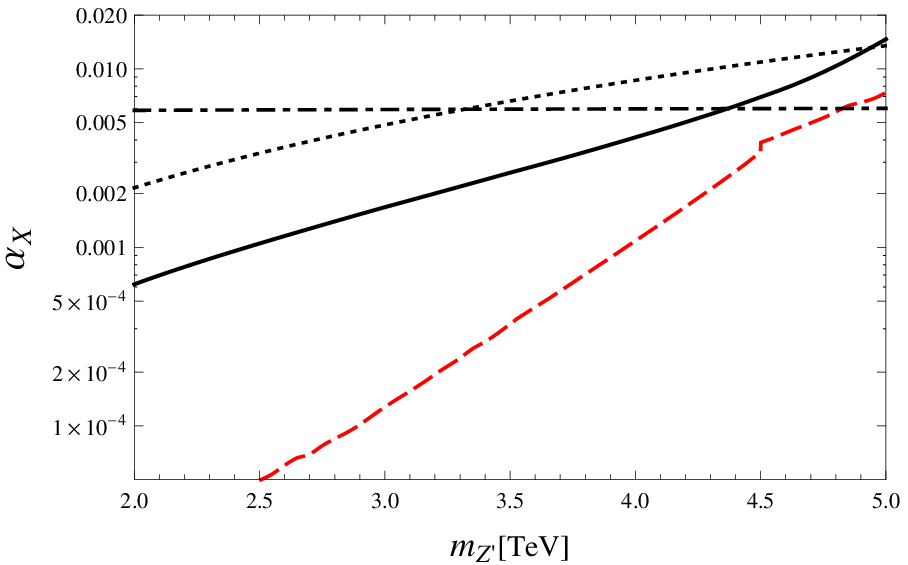}
\end{center}
\caption{
Allowed parameter region for the $Z^\prime$-portal RHN DM scenario. 
The top-left panel shows the results for the minimal $B-L$ model limit ($x_H=0$). 
The (black) solid line denotes the lower bound on $\alpha_X$ as a function of $m_{Z^\prime}$ 
    to reproduce the observed  DM relic abundance. 
The lower dashed line (in red) shows the upper bound on $\alpha_X$ 
   obtained from the search results for $Z^\prime$ boson resonance at the LHC. 
The shaded region is the final result after combining the cosmological and the LHC constraints, 
  leading to the lower mass bound of $m_{Z^\prime} \gtrsim 3.6$ TeV. 
For a comparison, we have also shown the upper long-dashed line (in red) 
  obtained in Ref.~\cite{Okada:2016gsh} by using the LHC results in 2015.   
The LEP upper bound in Eq.~(\ref{LEP}) is depicted as the dotted line. 
We also show the perturbativity bound on $\alpha_X$ as the dashed-dotted line. 
The top-right, the bottom-left and the bottom-right panels are same as the top-left panel, 
  but $x_H=-1$, $-2$ and $+1$, respectively.  
}
\label{fig:Final-xH}
\end{figure}

\section{Complementarity between the cosmological and the LHC constraints}
\label{sec:5}
Now we combine the constraints that we have obtained in the previous two sections. 
The RHN DM abundance has led to the lower bound 
   on the U(1)$_X$ gauge coupling for fixed $m_{Z^\prime}$ and $x_H$, 
   while the upper limit on the production cross section of the $Z^\prime$ boson at the LHC
   has derived the upper bound on the gauge coupling. 
Therefore, the two constraints are complementary to each other and, once combined, 
   the model parameter space is more severely constrained.

We show the results for various $x_H$ values in Fig.~\ref{fig:Final-xH}. 
The top-left panel shows the results for the minimal $B-L$ model limit ($x_H=0$). 
The (black) solid line shows the lower bound on $\alpha_X$ as a function of $m_{Z^\prime}$ 
    to reproduce the observed DM relic abundance. 
The lower dashed line (in red) shows the upper bound on $\alpha_X$ 
   obtained from the search results for $Z^\prime$ boson resonance 
   by the ATLAS~\cite{ATLAS:2016} and the CMS~\cite{CMS:2016} collaborations. 
Here, the ATLAS and the CMS bounds are combined as in the right panel on Fig.~\ref{fig:ATL-CMS}. 
The shaded region is the final result after combining the cosmological and the LHC constraints, 
  leading to the lower mass bound of $m_{Z^\prime} \gtrsim 3.6$ TeV. 
For a comparison, we have also shown the upper long-dashed line (in red), 
  which is obtained in Ref.~\cite{Okada:2016gsh} 
   from the ATLAS~\cite{ATLAS:2015} and the CMS~\cite{CMS:2015} results with the 2015 data.  
We can see the dramatic improvement from the previous result of  
   $m_{Z^\prime} \gtrsim 2.5$ TeV.   
The upper bound on $\alpha_X$ from the LEP constraint in Eq.~(\ref{LEP}) is depicted as the dotted line, 
   which turns out to be weaker than the LHC bound. 
We also show the theoretical upper bound on $\alpha_X$ in Eq.~(\ref{pert}) as the dashed-dotted line. 
If we impose this bound, it provides the most severe upper bound 
  for the range of $4.5$ TeV $\lesssim m_{Z^\prime} \lesssim5.0$ TeV.   
In Fig.~\ref{fig:Final-xH}, the top-right, the bottom-left and the bottom-right panels 
  are same as the top-left panel, but $x_H=-1$, $-2$ and $+1$, respectively.  
We find that the largest allowed region is obtained for $x_H \simeq -1$, 
  while no allowed region has been found for a $x_H$ value outside the range of $-2.5 \leq x_H \leq 1$.  
  
\begin{figure}[t]
\begin{center}
\includegraphics[width=0.6\textwidth,angle=0,scale=1.06]{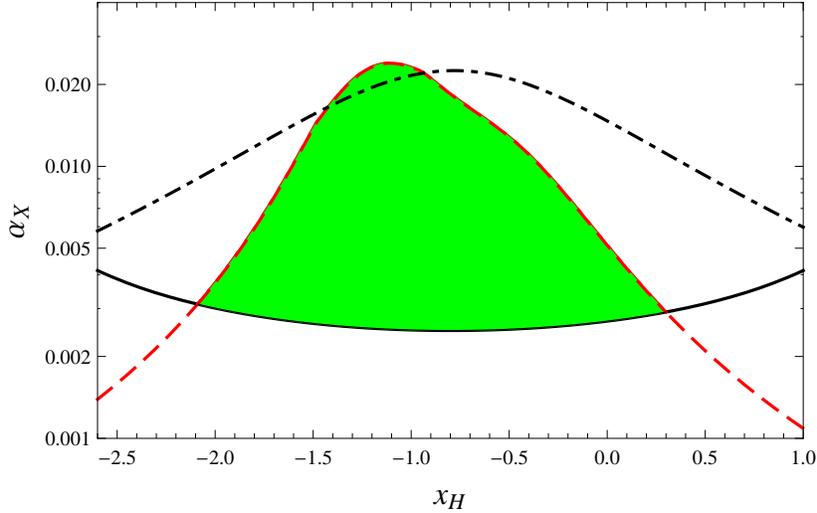} 
\end{center}
\caption{
Allowed parameter region for the $Z^\prime$-portal RHN DM scenario 
  for $m_{Z^\prime}=4$ TeV. 
The (black) solid line shows the cosmological lower bound on $\alpha_X$ as a function of $x_H$. 
The dashed line (in red) shows the upper bound on $\alpha_X$ 
   obtained from the combined ATLAS and CMS bounds.    
The shaded region is the final result for the allowed parameter space 
   after combining the cosmological and the LHC constraints, 
   leading to the allowed range of $-2.1 \leq x_H \leq 0.3$. 
The LEP bound appears above the plot range. 
The dashed-dotted line denotes the theoretical upper bound on $\alpha_X$ in Eq.~(\ref{pert}). 
}
\label{fig:4TeV}
\end{figure}

Finally, for a fixed $m_{Z^\prime}=4$ TeV, we show the allowed parameter region 
  in Fig.~\ref{fig:4TeV}. 
The (black) solid line shows the lower bound on $\alpha_X$ as a function of $x_H$ 
    to reproduce the observed DM relic abundance.  
As discussed in Sec.~\ref{sec:3}, the minimum $\alpha_X$ appears at $x_H \simeq-0.8$.    
The dashed line (in red) shows the upper bound on $\alpha_X$ 
   obtained from the combined ATLAS and CMS constraints. 
The shaded region is the final result for the allowed parameter space 
   after combining the cosmological and the LHC constraints, 
   leading to the allowed range of $-2.1 \leq x_H \leq 0.3$. 
The LEP upper bound appears above the plot range. 
The dashed-dotted line denotes the theoretical upper bound 
   from the perturbativity of the running $\alpha_X(\mu)$ up to the Planck scale.

The maximum value of $\alpha_X$ to satisfy the LHC bound appears at $x_H \simeq -1$. 
This means that the cross section of the $Z^\prime$ boson production at the LHC 
    exhibits its minimum at $x_H \simeq -1$. 
This fact can be roughly understood by using the narrow width approximation. 
When the decay width of the $Z^\prime$ boson is very narrow, 
   we approximate Eq.~(\ref{CrossLHC2}) as 
\bea 
{\hat \sigma}(q \bar{q} \to Z^\prime \to  \ell^+ \ell^-) \simeq 
\frac{\pi}{1296} \alpha_X^2
M_{\ell \ell}^2 
\left[ \frac{\pi}{m_{Z^\prime} \Gamma_{Z^\prime}} 
\delta( M_{\ell \ell}^2-m_{Z^\prime}^2)
\right]
F_{q \ell}(x_H) 
\propto \frac{F_{q \ell}(x_H)}{F(x_H)}.  
\eea   
Using the explicit formulas for $F(x_H)$ and $F_{q \ell}(x_H)$ given in Eqs.~(\ref{F}) and (\ref{Fql}), 
     we can verify that the function $F_{q \ell}(x_H)/F(x_H)$ exhibits a minimum at $x_H\simeq -1$.

\section{Conclusions}
We have considered the minimal non-exotic U(1)$_X$ extension of the SM, 
    which is free from all the gauge and the gravitational anomalies 
    in the presence of three right-handed neutrinos. 
After the breaking of the U(1)$_X$  and the electroweak gauge symmetries, 
    the SM neutrino masses and flavor mixings are generated through the seesaw mechanism.
We have extended this model by introducing a $Z_2$-parity and assigned an odd-parity to 
    one RHN while even-parities to all the other particles.  
Thanks to the parity, the $Z_2$-odd RHN is stable and hence the DM candidate. 
No extension of the minimal particle content is necessary to incorporate a DM candidate into the model. 
With the other two RHNs, the seesaw mechanism works 
   to account for the neutrino oscillation data with one massless neutrino. 
In this model, the RHN DM communicates with the SM particles through the $Z^\prime$ boson exchange. 
We have investigated this $Z^\prime$-portal RHN DM scenario in this paper.

Phenomenology of the scenario is controlled by only four free parameters, namely, 
   the U(1)$_X$ gauge coupling ($\alpha_X$), the RHN DM mass ($m_{DM}$), 
   the $Z^\prime$ boson mass ($m_{Z^\prime}$) and the U(1)$_X$ charge of 
   the SM Higgs doublet field ($x_H$).  
We have first considered the cosmological constraint of the scenario.  
In order to reproduce the observed DM relic density, we have found it necessary 
   to enhance the DM annihilation cross section via $Z^\prime$ boson resonance.  
Therefore, the RHN DM mass is always set to be $m_{DM} \simeq m_{Z^\prime}/2$, 
   and the number of the free parameters is reduced to three. 
The three parameters are constrained by the DM relic abundance. 
For example, the lower bound on $\alpha_X$ has been obtained 
  as a function of $m_{Z^\prime}$ for a fixed $x_H$. 
We have next considered the LHC constraints on the $Z^\prime$ boson production cross section 
   by employing the most recent results by the ATLAS and the CMS collaborations 
   on the search for a narrow resonance with the di-lepton final state.  
We have derived  the lower bound on $\alpha_X$  as a function of $m_{Z^\prime}$ for a fixed $x_H$. 
In constraining the model parameter space, the cosmological and the LHC bounds 
   are complementary with each other, and we have narrowed the phenomenologically 
   viable parameter region by combining them. 
For example, we have found the lower limit of the $Z^\prime$ boson mass to be $m_{Z^\prime} \gtrsim 2.7$ TeV.   
In our analysis, we have also taken into account other phenomenological constraints such as 
   the LEP bound on the U(1)$_X$ symmetry breaking scale and 
   the perturbativity bound on the running U(1)$_X$ gauge coupling below the Planck scale.

\section*{Acknowledgments}
We would like to thank Ryusuke Endo for valuable discussions and comments. 
We also wish to thank Digesh Raut for reading the manuscript and his useful comments. 
S.O. is very grateful to Andy Okada for his encouragements.  
She would also like to thank the Department of Physics and Astronomy at the University of Alabama
  for hospitality during her visit for the completion of this work. 
The work of N.O. is supported in part by the United States Department of Energy (Award No.~DE-SC0013680). 

\bibliographystyle{utphysII} 
%
{
\def\section*#1{}
\bibliography{./bibliography}                           
}

\end{document}